# Different but Equal: Comparing User Collaboration with Digital Personal Assistants vs. Teams of Expert Agents

Claudio S. Pinhanez, Heloisa Candello, Mauro C. Pichiliani,
Marisa Vasconcelos, Melina Guerra, Maíra G. de Bayser, Paulo Cavalin
IBM Research – Brazil
csantosp@br.ibm.com

This work compares user collaboration with conversational personal assistants vs. teams of expert chatbots. Two studies were performed to investigate whether each approach affects accomplishment of tasks and collaboration costs. Participants interacted with two equivalent financial advice chatbot systems, one composed of a single conversational adviser and the other based on a team of four experts chatbots. Results indicated that users had different forms of experiences but were equally able to achieve their goals. Contrary to the expected, there were evidences that in the teamwork situation that users were more able to predict agent behavior better and did not have an overhead to maintain common ground, indicating similar collaboration costs. The results point towards the feasibility of either of the two approaches for user collaboration with conversational agents.

## 1 INTRODUCTION

The vast majority of the conversational systems available today follow a one-to-one paradigm where the user interacts either with *personal assistants*, such as *Apple*'s *Siri* or *Amazon*'s *Alexa*, or with *expert agents* provided by enterprises [26, 47]. We explore in this paper an alternative model where users accomplish goals by collaborating with a *team of expert agents* with in the context of a shared chat. Figure 1 depicts the architecture and the basic messaging mechanisms of the three models.

Suppose the user's goal is to organize a trip to a touristic destination. In the *expert agent* model, the user can converse individually with the providers of flights, hotels, and attractions, collect the information, and manually coordinate with each of them. Alternatively, she can use a *personal assistant* which communicates, under the covers, with the providers of those information and services, negotiates with them, and brings the results to her. All access to user information and conversation data is controlled by the personal assistant which works as an artificial butler. Finally, she can summon a *team of expert agents* (and possibly her own personal assistant) into a common, shared chat in which the expert chatbots talk directly to her, listen to each other, and collaborate to organize the trip as if they were in a meeting.

Using the expert agents model normally means that the burden of coordinating the different service providers is carried solely by the user, since the agents are unaware of each other. We do not study this model in this paper because we want to address the other two models in which some of the coordination tasks are offloaded from the user. There are advantages and drawbacks in using the personal assistant or the team of expert agents models, including technical constraints and challenges, user preferences, and even ethical and moral considerations. Although we present and discuss briefly those issues in this paper, the key goal of this work is to explore empirically how a team of expert agents is different from a personal assistant in terms of accomplishing tasks and facilitating the teamwork between the user and the agents.

Based on the human experience, we would expect that teamworking with a group of agents to be more difficult than using a single personal assistant but would potentially yield better outcomes. But by doing comparative studies with two equivalent chatbot systems, one similar to a personal assistant and the other configured as a conversation with a team of expert chatbots, we have found evidences not only that participants seem to be able to accomplish their goals equally well in both contexts but also that participants seem to have a different kind of collaboration process with the team of expert agents compared to when interacting with the personal assistant. We perform the collaborative part of this analysis using Clark's *joint activity* [11] as our theoretical framework, as adapted to *joint human-agent activity* by Klein et al. [31].

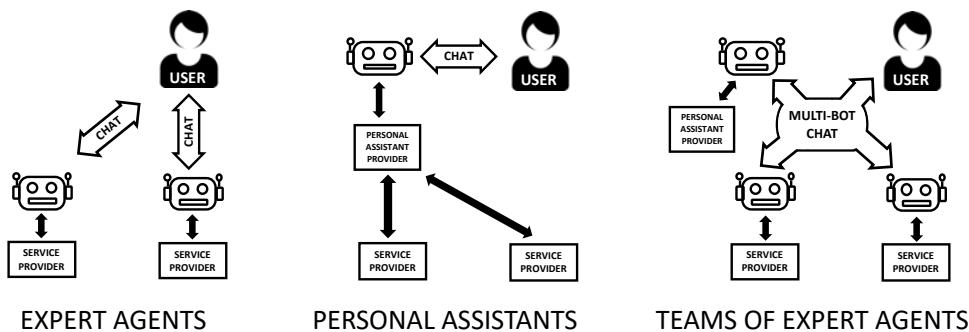

Figure 1: Three models of conversational user interaction with service providers.

Our empirical investigation was done in two studies, described in this paper, which exposed participants to chatbots which provide financial advice about low-risk investments in a conversation format similar to popular chat apps, such as *WhatsApp*. Using as a starting point a finance advice multi-expert system which employs four different chatbots called *finch* [17], we created a personal assistant version by simply remaking the user interface so it appears that all machine utterances are coming from the same chatbot. Therefore, by construction, both versions produced identical utterances and behavior.

The paper starts by reviewing research and practice in multi-bot conversational systems and the main technical challenges to build them. As an example, we describe the financial advice system used in the studies and how some technical challenges in building multi-bot systems, such as turn-taking management, were handled. Following, we introduce the theoretical frameworks used in the paper, related works, and our fundamental research questions. We then describe each of the two studies and their main findings, followed by discussions about how they address our research questions. We end the paper by reviewing and discussing our results, exploring the limitations of our studies, and considering the implications of our results for the choice between personal assistants and teams of expert agents.

## 2  MULTI-AGENT CONVERSATIONAL SYSTEMS

For the purposes of this paper, we consider a *conversational system* any machine system which interacts with its users primarily through conversation, with structures and conventions of natural human language (or some reasonable resemblance of). A *multi-agent conversational system* is a conversational system in which the user interacts with more than one computational agent at the same time. We use the term *chatbot* to specifically refer to conversational systems which engage in dialogue with their users by producing and recognizing complex language sentences, mostly to contrast with old-fashioned *interactive voice response (IVR)* systems [14]. We employ the term *(digital) personal assistant* to refer to a conversational system which takes the role of assisting the user, often by intermediation, in accomplishing everyday personal or professional tasks. Also, an *expert agent* is considered here as a conversational system which interacts with a customer on behalf of a company or government agency. A *team of expert agents* consists of a group of expert agents or personal assistants which interact with the user and to each other, using natural language in the context of a shared conversation. In most of the paper, we narrow those terms down to *personal chatbots, expert chatbots, and multi-bot chats,* respectively, to highlight the use of a rich text-based chat context.

The rapid progress of natural language processing in the recent years, fueled by notable advances in *Artificial Intelligence* and *Neural Networks*, have made such conversational systems not only possible, but a reality available at the fingertips of any smartphone user. However, in many ways the technology used in most commercial deployments today is not much different from the one of the early chatbots [59], based on the use of *intention-action* API services in which answers are manually written, one-by-one. Examples of those API services are *Facebook's wit.ai*, *IBM's Watson Assistant, Microsoft's LUIS, Amazon's Lex,* and *Google's Conversation APIs*, as well as an increasing number of SDKs provided by startups and open source initiatives. Clearly, the intention-action paradigm poses significant limits on what a chatbot can do since it is limited by the amount of text or code created by its developers. Also, conversational systems built in this way require many examples of each possible utterance of the users, resulting often in failures to understand user utterances and brittleness when asked about subjects and concepts beyond its competence.

Moreover, most of the current conversational systems are structured as the conversation of one user with one single agent. The ultimate goal of our work is to inform the design of multi-bot chat systems, in which users can successfully converse with a team of expert agents. Human collaboration with multiple agents has been studied before, notably in the context of managing swarms of robots [1, 8, 27, 43, 45], disaster response [51], and healthcare collaborative systems [3, 20, 28, 29]. More recently, there has been interest in *human-agent collectives* [27, 51] in which large numbers of users have their activities coordinated by automated agents. However, there are very few examples of actual systems in which the teams of expert agents use natural language to collaborate [16, 23, 44] although there has been some theoretical research in the area [2, 10, 16, 30, 61]. Of course, before the recent emergence of chatbot-building API services and toolkits, part of the problem was just the difficulty of developing each chatbot itself. But it is likely that the scarcity of actual multi-bot chats is related that they pose some technical challenges to developers as discussed next.

## 2.1 Technical Challenges of Building Multi-Agent Conversational Systems

A key challenge in creating expert agents which can work together using a shared chat is *turn-taking* [57], that is, how the agents in the conversation coordinate their utterances so each of them participates at the appropriate times. Linguistics has studied turn-taking in humans, notably by Sacks et al. [54], and, in broad terms, has found that participants in a conversation normally do not overlap their utterances. In everyday conversations, the content of the last utterance tends to make clear to all the participants who should be the *selected-next-speaker (SNS)* [54]. For instance, the current speaker can direct who should be the next by using some form of *direct address* [12], such as a vocative with the name of the next speaker which is uttered to designate the SNS. Technology for managing turn-taking in multi-bot chats is still in its infancy, and in our studies, we used an agent orchestration tool called *Ravel* [17], described briefly in the following subsection.

There has been a lot of recent work on orchestrating groups of hidden chatbots as a way to create a single chatbot, notably as part of the *Amazon's Alexa prize* of 2017 [50] to build a chatbot which could manage to chitchat for 20 minutes with users. Some of the best contestants used architectures where multiple chatbots with different skills received each user utterance, produced an answer, and an arbitration scheme decided which of the multiple answers was the best [46, 55]. Similarly, a system called *Evorus* [24] used not only multiple chatbots but also crowd-sourced human workers when it was unable to answer the user.

Other technical challenges facing the developers of multi-bot chats include the need of managing the access and distribution of the utterances among the agents, for instance using a communication hub [17]. Interestingly, in computer-mediated conversation of human beings, such as in chats, overlapping of utterances is also a problem: unrelated messages and threads happen in a likelihood proportional to the number of active participants involved in the communication [22]. Finally, testing multi-agent systems is always more complicated than working with single chatbot systems, often requiring the employment of user simulators [58].

## 2.2 *finch*: A Multi-Bot Investment Advice System

In the two studies described in this paper, we used a multi-bot chat system called *finch* which provides, in Portuguese language, information and decision support on low-risk investments. Following a user-centered design process focused on middle-income users with little knowledge of finances, *finch* was designed as a multi-bot chat where a user can converse with four chatbots. Three of the chatbots are expert agents in typical low-risk Brazilian financial products: *PoupançaGuru (P),* expert in savings, *TDGuru(TD)*, expert in treasury bonds, and *CDBGuru (CDB)*, expert in government bonds. There is also *InvestmentGuru(In)*, a kind of personal adviser which helps to summarize the conversation and prompts users to make more questions (similar to what was proposed by Bradshaw et al. in [8]). The chatbots were built using *IBM Watson Conversation Services* APIs. The interface to the system resembles a *WhatsApp* chat (see fig. 2).

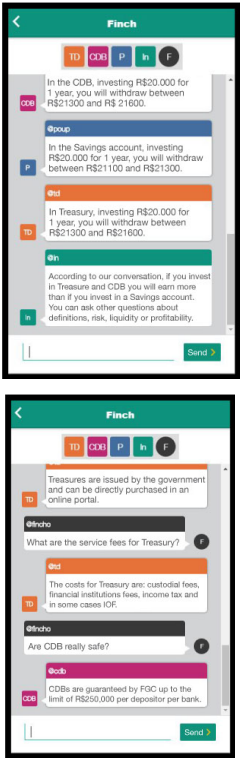

Figure 2: Screenshots of *finch*.

| | |
|---|---|
| In | Hello, I am your investment adviser. […introductions by gurus…] Our gurus can help with information about Savings, CDB, and Treasury Bonds. Type the amount in reais (R$) and the duration of the investment. |
| **User** | **R$5000 2 years** |
| In | @cdbGuru @poupancaGuru @tesouroGuru, what would be 5.000 reais after 2 years? |
| CDB | In CDB, if R$ 5.000 were invested for 2 years, you would get between R$ 5700 and R$ 5800 |
| TD | In Treasury Bonds, if R$ 5.000 were invested for 2 years, you would get between R$ 5700 and R$ 5800 |
| P | In Savings, if R$ 5.000 were invested for 2 years, you would get between R$ 5500 and R$ 5600 |
| In | Based on our conversation, CDB and Treasury Bonds yield more in this time frame. You can make other questions about definitions, risk, liquidity, or rentability to our gurus. |
| **User** | **What is CBD?** |
| CDB | CDB (Certificado de Depósito Bancário) works as a loan which you make to the bank, for which you receive remuneration. |
| **User** | **Which investment has more risk?** |
| In | To make direct questions to our gurus, type @ followed by the name of the guru (poupançaGuru, cdbGuru, or tesouroGuru) |
| **User** | **Which are the risks of CDB?** |
| CDB | If your investment is up 250 thousand reais, it is covered by the Fundo Garantidor de Crédito. |
| **User** | **With Treasury Bonds, do I pay taxes?** |
| TD | The costs of Treasury Bonds are: deposit fee, banking fee, income tax, and in certain cases IOF. |

Figure 3: Partial transcript of a user interacting with *finch*, translated from the original in Portuguese.

To implement turn-taking, we used a domain specific language for generation of conversation rules developed by Bayser et al. [17], which was implemented in a conversation governance system called *Ravel*. [17]. In *finch*, turn-taking is accomplished by considering multiple characteristics of the chatbots. The investment counselor chatbot (*InvestmentGuru*) mediates the conversation with the user, while the investment products experts reply only when required. The expert chatbots answer either when they are addressed directly or when the user's utterance topic is about their expertise. By default, the last chatbot which spoke is the one that answers the next user utterance, except in the case where the user sends utterances related to simulation of the return of investment which are always mediated by *InvestmentGuru*. The *finch* chatbots also take in account the kind of speech act uttered by the last speaker, user or chatbot. A *question* is often followed by an *answer*, a *greeting* by a *greeting*. All *finch* experts were designed to reply to *greetings, thank, query, request, inform, agree, not understood,* and *bye* speech acts, but they do so only when the user utterance is classified to be on their topic of expertise. A typical conversation with a user is depicted in the partial transcript shown in fig. 3.

The *finch* system was created and designed for Brazilian bank customers with very limited knowledge of finances, typically young, middle-class, technically-enabled users who wanted to make quick decisions to invest small amounts of money for short times. In Brazil, due to the high inflation, there are multiple low-risk, short-term financial products which yield significant but different returns depending on the economic conditions and on specific bank offerings, among them the three which were represented by the expert chatbots. Notice that although the three products are risk-free (the customer will always receive at the least what was invested) the choice of investment is not trivial given the different variables which affect the country's economy and the rates offered by each bank in the market. For instance, a typical 3-year treasury bond in 2016 was returning 39% yearly compared to 29% in savings, while in 2018 there was almost no difference, making the choice of saving accounts a better option given its daily liquidity. Choosing the right low-risk investment can be a quite complicated decision-making process which involves gathering the offers available and understanding their expected returns, liquidity, and tax impacts.

# 3 A PERSONAL CHATBOT OR A MULTI-BOT CHAT?

At first look, it seems that getting things done by interacting with multiple chatbots is an unnecessary increase in complexity for our lives. If a single personal assistant can contact directly all the services we need to make a decision or accomplish a task, why it would be better to move to a scenario which looks like adding yet another "meeting" to our over-stressed schedules?

Previous research on the user experience of single- vs. multi-bot conversational systems have produced mixed results. Maglio et al. [39], in a *Wizard-of-Oz* study exploring speech interaction in an office setting, found that participants seem to have less effort controlling multiple devices with a single conversational agent instead of talking to each device directly. In another *Wizard-of-Oz* study, Chaves and Gerosa [10] found no significant difference in the conversation structure, but participants reported more confusion using a multi-bot chat [10].

However, the design process of *finch* and the ways some of its first users worked with it, suggested us that multi-bot chats could be effective in unusual ways which have not been properly studied. This led us to the main focus of this work, which is to determine whether a team of chatbots is more difficult to work with than with an equivalent personal chatbot. In particular, we wanted to explore whether collaborating with a team of chatbots would affect the final decision made by the users, and whether a likely increase in collaboration and coordination costs (commonly observed when teamworking with people) would also be observed in such context. To properly address those issues, it is first necessary to review some of the related work and theory, particularly in the context of collaboration of people and machines.

## 3.1 Theoretical Background

Many factors may influence whether working with multi-bot chat instead of a personal chatbot is easier or more difficult, both in perceptual and in effective terms. To focus our research, we address in this paper two main factors, (1) whether one of the options is better in facilitating the achievement of the user's task, and (2) whether working in teams of machines instead of a single machine makes the collaboration process different, possibly increasing collaboration costs.

**Task Outcomes and Expertise**

Starting with the former, the studies described in this paper test whether and how the task outcome is affected if the user interacts with a single chatbot vs. a team of chatbots. Since our experiments, as described later, are related to supporting the decision-making process, we choose to examine deeply whether the choice between the two models affects the object of the decision. Since often expert decision-making is affected by the user's perception of the system's expertise, we selected three key attributes often associated with human expertise, *competence*, *effectiveness*, and *trustworthiness,* and explored whether users perceive those attributes differently after interacting with a single machine expert vs. a team of machine experts.

Conversational systems are more often seen in failure situations than conventional computer systems and, as therefore, tend to generate higher levels of doubt in their *competence*. Moreover, there is evidence that when users are exposed to situations where they see machines failing, they tend to lose their confidence on their reliability faster than if a human being was failing at the same rate, in a phenomenon called *algorithmic aversion* [15]. The perception of the competence of a chatbot and how to manage it adequately are, therefore, very important issues that designers and developers of chatbots must address.

The concept of *effectiveness* of a chatbot is deeply linked to traditional questions of usability. However, in conversational systems, the difficulties are often aggravated by two factors: there are not many structural clues about how to do things (such as menus, forms); and language recognition mistakes make the navigation process harder. Here, we are not concerned whether a user accomplishes a task or not (which will be determined in other ways), but rather if she believes she knows which actions to perform to get her task done with the conversational system.

Even when a user finds a conversational system competent and helpful in a domain, she may still not *trust* it, and this may significantly affect the decision-making process. A key differentiating element is that conversational systems are often anthropomorphized by their users [30], and therefore perceived as having their own motivations and goals [32]. For instance, a financial advice chatbot being provided by a bank may be perceived as trying to maximize bank's profits at the expense of the client's.

**Coordination Costs**

To address the second main factor considered in our study, *collaboration costs*, we choose as a theoretical foundation for this work Clark's concept of *joint action*, as "…one that is carried out by an ensemble of people acting in coordination with each other." [11], p.3. Joint actions are often part of *joint activities* which comprise both individual and joint actions by a group of people working towards goals in socially constructed contexts. The realization of joint activities requires the involved parties to coordinate both the individual and joint actions, and the process in which the activity is accomplished. The basic source of information to support the coordination of activities is referred as *common ground*, which is the set of all knowledge, beliefs, and assumptions shared by the participants about the joint activity (see [11], chapter 4). Maintaining and repairing common ground is a key part of joint activity in which language plays is an essential part. For Clark, it is almost as language is collaboration and vice-versa.

Many authors have looked whether human-human collaboration theories are applicable to human-agent collaboration [7, 8, 33]. In particular, Kramer et al. [33] argued that since users tend to behave with and treat computers as human beings (as pointed, among others, by Reeve and Nass [53]), "…due to our social nature humans will use their interaction routines also when confronted with artificial entities." In the particular case of joint activity, Klein et al. [31] regarded human-agent collaboration to fall within the context of socio-technical systems, and identified four basic requirements for their joint activity, that the parties (1) enter into an *agreement*; (2) be mutually *predictable*; (3) be mutually *directable*; and (4) maintain *common ground*.

Agreement is a pre-condition to collaboration in the sense that parties must concur in who the participants are, what processes must be followed, and what accomplishing a goal is. Mutual predictability is required to allow easier coordination: one party can expect that some action is going to be performed by another in some conditions. Mutual directability assures that one party may ask another to perform an action or a goal during the joint activity. Common ground includes not only the pertinent knowledge, beliefs, and assumptions among parties [32], but also the context of the joint activity. Importantly, Klein et al. [31] considered situations where at least one of the team players is a machine, called *joint human-agent joint activities*, and argued that the four basic requirements also apply to them.

Current conversational systems are very limited in meeting the requirements for effective teamwork as propose by Klein et al. [32], and leave the brunt of the coordination task to the user. Therefore, we focus this paper on understanding the user experience of joint human-agent activities in those systems. It has been shown that there are additional coordination costs in teamworking with multiple chatbots arising from the complexity of conversing with multiple parties. For instance, the conversational experience with one human being is different with that of talking with multiple people [6, 22, 57], albeit the foci of those works was the resulting linguistic structure. In the realm of joint activities with multiple agents, the studies performed so far have also dealt primarily either with the impact on task accomplishment [34, 35] or with linguistic effort [10, 39]. We are unaware of any research work which have looked into conversational collaboration with machines under the theoretical lens of joint-activity theory.

## 3.2 Research Questions

Is a team of expert chatbots more difficult or easy to work with than with an equivalent personal chatbot? Although answering this question is the final aim of this work, we have first to consider that "difficulty" is an ill-defined concept in this context, rendering the question almost meaningless as an object of scientific study. Based on the related work just described, we propose breaking this question into two simpler, easier to address questions, focusing first on how working with a team of chatbots may affect the outcomes of the collaboration task; and second, on how it may affect the process of collaboration. This is formally captured in the following two research questions:

RQ1: Is the **task outcome** of collaborating with a personal chatbot or an equivalent multi-bot chat similar?

RQ2: Is the **process** of collaborating with a personal chatbot or an equivalent multi-bot chat similar?

Beyond simply answering those two research questions based on results of experimental studies, we believe it is fundamental to understand the possible reasons and causes of those results. To better understand factors which could influence the task outcome (RQ1), we decided to look deeper into whether users have different perceptions of the expertise level for each model. Based on our previous discussion on expertise, we examine further whether users of a personal chatbot or an equivalent multi-bot chat perceive their **competence** as similar, that is, whether each system seems to have the knowledge needed to help them achieve their task. Similarly, we studied the users' perception of **effectiveness**, understood here as the ability of the system to actually support the task. And finally, we thought it would be important, given the financial context, to verify whether the two different models produced different perceptions of **trustworthiness**.

Furthermore, to better explore the second research question (RQ2), we consider the framework for joint human-agent activity discussed before. It is important to see that the user is a team player in all cases and, as discussed, the main responsible for most of the coordination tasks such as common ground maintenance and repair. If the burden for the user to accomplish her task in a multi-bot chat increases too much due to larger collaboration costs, it is unlikely that users will opt for multi-bot chats instead of personal chatbots.

Given the framework for human-agent joint-activities proposed by Klein et al. [31, 32], we look into whether each different chat model affects the user's ability to **predict** or **direct the behavior** of the agents she is collaborating with; and whether the user's is able to **manage common ground** as effectively in each case. In our particular experimental scenario, **agreement** issues were virtually identical in both conditions.

### 3.3 Comparing the Collaboration with Personal Chatbots vs. Multi-Bot Chats

With the aim to explore our research questions we designed two studies, in which two versions of the *finch* system were created to be evaluated by the participants. In the first version, a single financial adviser, similar to a personal assistant, conversed with the participants. In the second version, we used the original version of *finch* where a team of expert chatbots collaborates to answer the questions from the user.

For simplicity, we denominate the two versions of *finch* used in the study as *single-bot* and *multi-bot*, respectively, though keeping in mind that they are, in fact, representatives of the personal chatbot and the multi-bot chat models. Both versions were built by simply changing the way the content was delivered on the chat interface. In the single-bot version, all the information was delivered as if it was produced by the *InvestmentGuru* chatbot, even though some utterances are in fact produced by other three (hidden) chatbots. The two versions were constructed this way to assure that the participants of the single- and the multi-bot versions would have essentially the same experience, not only in terms of the corpora of knowledge but also the actual text and conversational structure.

Perceptually, the main difference between the single- and the multi-bot experiences was what the user would see in the interface. In the single-bot version, all the utterances from the system were shown coming from the *InvestmentGuru (In)* green chatbot (see fig. 4). The multi-bot version had red, purple, and blue chat bubble headings to represent each expert bot and green for *InvestmentGuru* (see fig. 2). The participant's utterances are shown in black heading bubbles which appeared from her icon on the right. Also, in the multi-bot version participants could address directly the expert chatbots by using the vocative @*chatbot_name*. Importantly, participants of the multi-bot version were not told that they were working with multiple chatbots, they had to figure out directly from the representation. Notice that the interface replicated common ways of displaying multiple users in popular chat applications such as *WhatsApp*.

## 4 THE FIRST STUDY: BETWEEN-SUBJECTS UNSUPERVISED INTERACTION

The first study consisted of a between-subjects experiment run without human supervision in which participants answered a demographics questionnaire, then interacted with just one of the two versions of *finch*, and finished with a survey about their experience. No reward or money incentives were given to participants. The data collected in the questionnaires as well as in the logged conversations was then analyzed to produce the study findings.

Participants did not use real money for two reasons. First, we wanted to frame the study according to each participant economical context [36], in which each participant followed a personally relevant scenario as explained later. Second, the three investment products are zero-risk investments tailored for conservative investors, and this profile was also a criterion to participate in the study. Choosing among them is more an issue of information gathering than of weighing risks, so typical risk-avoiding behaviors are not present in our experiment. Therefore, we deemed unnecessary to counterbalance risk-avoidance with a real money context or economic rewards.

## 4.1 Methodology of the First Study

The procedure consisted of 5 phases: invitation, introduction and disclaimer, demographics and knowledge questionnaire, free interaction, and evaluation questionnaire. All participants had to be fluent in Portuguese.

**Invitation:** participants were initially recruited by a snowball sample under two conditions: more than 18 years old; well-educated (university student or higher). They received an invitation by e-mail to evaluate a financial adviser system with a web link. The link randomly sent them to either the single- or the multi-bot version. The procedure was identical for each version. No reward was given to them to go through the study.

**Introduction and disclaimer:** participants were welcomed with a brief explanation (the same as in the e-mail invitation). They were also guaranteed the secrecy of any information provided. Next, the participant wrote his or her name and agreed with a disclaimer.

**Demographics and financial knowledge questionnaire:** the aim of the first survey was to assess participants' expertise in financial investments, technology expertise, and demographics. Next, the participant read a scenario framing the interaction: *Imagine you received the equivalent of your monthly salary as a cash prize this month. You would like to invest this prize and your plan is not to use this money for 2 years. You need help deciding what type of investment is the best for you. For that you will use FINCH, a financial advice app.*

**Free interaction:** as mentioned before, participants were randomly assigned to one of the two versions. The single-bot version started with *InvestmentGuru* presenting itself as a financial adviser which could provide information about savings, CDB, and treasury bonds, and asking the participant for an amount and a duration of investment. The multi-bot version started with *InvestmentGuru* also presenting itself as a financial adviser but in this case, it was followed by the expert chatbots introducing themselves as experts in their corresponding financial products. After the introductions, *InvestmentGuru* asked the participant for an amount and a duration of investment. In both versions, participants were free to make any questions or comments. The *finch* system answered the questions and utterances by determining their intent and allocating the proper chatbot(s) to respond. If requested, both systems could simulate the return of the investments and make simple comparisons among the three options.

**Evaluation questionnaire:** in the two versions, the evaluation questionnaire was made available to users either after 5 minutes of inactivity or when they typed utterances such as "thanks", "bye", or "I decided to invest". In both cases, participants then answered the following questions (translated from Portuguese). All questions except Q1 were answered by choosing among *disagree*, *partially disagree*, *indifferent*, *partially agree*, *agree*.

- Q1. In which investment option would you invest? (choose one option)
  Savings : CDB : Treasury Bonds : Indifferent : None
- Q2. This system would help me to make a financial decision in a more informed way.
- Q3. This system has enough knowledge about finance to help me to take a decision.
- Q4. Considering what I learned now I could advice someone to make an investment.
- Q5. I would trust this system to make a decision.
- Q6. I would rather use this system than to speak with my bank account manager.
- Q7. I trust more this system than the manager of my bank.
- Q8. I reconsidered my current investment options after using this system.
- Q9. I would recommend this system to a friend.
- Q10. This system is innovative.
- Q11. This system does not look like any another financial advice system I used before.

After participants had filled in the evaluation questionnaire, the study finished with an acknowledgment message. All questionnaire and conversation data were logged in an *Elastic Search* database.

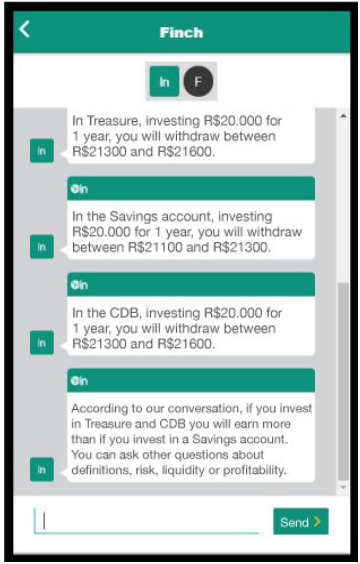

Figure 4: Single-bot version of *finch*.

Table 1: Results of financial knowledge questionnaire.

|  | SINGLE-BOT | | | MULTI-BOT | | |
|---|---|---|---|---|---|---|
|  | Low | Med | High | Low | Med | High |
| know savings | 6 | 18 | 9 | 10 | 17 | 9 |
| know cdb | 19 | 11 | 3 | 22 | 9 | 5 |
| know tr. | 18 | 12 | 3 | 27 | 5 | 4 |
| Int. in finance | 5 | 17 | 11 | 7 | 19 | 10 |

Table 2: Number of conversations and utterances for participants and chatbots.

|  | SINGLE-BOT | | MULTI-BOT | | TOTAL |
|---|---|---|---|---|---|
|  | number | % | number | % | number |
| conversations | 33 | 48% | 36 | 52% | 69 |
| avg.duration/conv.(s) | 364 | — | 387 | — | 376 |
| utterances | 1047 | 52% | 956 | 48% | 2003 |
| avg. utt/conv. | 31.7 | — | 26.6 | — | 29.0 |
| particip. utterances | 407 | 53% | 362 | 47% | 769 |
| avg. participant | 12.3 | — | 10.1 | — | 11.1 |
| all bot utterances | 640 | 52% | 594 | 48% | 1234 |
| all bot utt/conv. | 19.4 | — | 16.5 | — | 17.9 |

### 4.2 Data Analysis

The analysis of the data yielded by the questionnaires used standard statistical tools and methods as discussed later. The log analysis of the conversations was based on manually assigning 10 types of user intent to each user utterance. The labelling was performed by a group of labelers and majority voting was used in cases of disagreement to classify each user utterance. The 10 types of *user intent* were : *initial simulation*, for utterances done in response to the simulation prompted in the beginning of the conversation by *InvestmentGuru*; *core - cdb*, *core - savings*, and *core – treasury bonds*, for utterances asking expertise in each of the respective investments; *simulation* and *comparison*, for utterances asking for simulation of an investment across the three financial products or for a comparison among them, respectively; *generic finance*, for utterances regarding basic information about finance such as what liquidity is; *other investments*, for utterances asking information about financial products beyond the competence of *finch*; *chitchat*, corresponding to utterances such as "hello", "thanks", etc.; and *conversational problems*, used to indicate situations where the user was dealing with problems resulting from the inability of the system to answer a question or to perform an operation.

To simplify the analysis, the 9 types of user intent (excluding *initial simulation* since it was identical for all participants) were grouped into 4 *basic categories*: *core*, comprising the intents related to the three core bots; *weighing*, comprising *simulation* and *comparison*; *other finance*, comprising *generic finance* and *other investments*; and *conversing*, joining *chitchat* and *conversation problems*.

### 4.3 Results of the First Study

After two weeks of the study, we had 99 participants of which 69 completed all four phases of the study, 33 doing the single-bot version and 36 in the multi-bot version.

**Comparison Between the Two Groups of Participants**

A standard statistical analysis of the demographics questionnaire showed a good balance in gender and age and, more importantly, that the single- and multi-bot groups of participants had statistically similar distributions. Also, we did not see any statistically significant differences in the distributions of financial knowledge between the single-bot and multi-bot participants, as depicted in Table 1. Moreover, as shown in Table 2, the two groups tended to have very similar length of conversations, number of utterances, and average number of utterances by the participant and the chatbots.

Table 3: Results of Q1, "In which investment option would you invest?" of the evaluation questionnaire.

| Q1 | SINGLE-BOT | | MULTI-BOT | | TOTAL | |
|---|---|---|---|---|---|---|
| | number | % | number | % | number | % |
| SAVINGS | 1 | 3% | 4 | 11% | 5 | 7% |
| CDB | 11 | 33% | 14 | 39% | 25 | 36% |
| TREA. BONDS | 15 | 46% | 13 | 36% | 28 | 41% |
| INDIFFERENT | 3 | 9% | 1 | 3% | 4 | 6% |
| NONE | 3 | 9% | 4 | 11% | 7 | 10% |

Table 4: Results of the Q2 to Q11 questions of the evaluation questionnaire.

| | SINGLE-BOT | | | MULTI-BOT | | | equal? |
|---|---|---|---|---|---|---|---|
| | agree | same | disagree | agree | same | disagree | p-value |
| Q2 | 24 | 2 | 7 | 23 | 7 | 6 | 0.847 |
| Q3 | 18 | 2 | 13 | 20 | 5 | 11 | 0.693 |
| Q4 | 10 | 2 | 21 | 11 | 6 | 19 | 0.616 |
| Q5 | 17 | 0 | 16 | 17 | 6 | 13 | 0.629 |
| Q6 | 11 | 3 | 19 | 13 | 2 | 21 | 0.859 |
| Q7 | 9 | 8 | 16 | 8 | 5 | 23 | 0.363 |
| Q8 | 10 | 10 | 13 | 15 | 9 | 12 | 0.453 |
| Q9 | 22 | 3 | 8 | 25 | 2 | 9 | 0.845 |
| Q10 | 23 | 3 | 7 | 28 | 4 | 4 | 0.305 |
| Q11 | 17 | 8 | 8 | 26 | 8 | 2 | 0.099 |

**Findings of the Analysis of Evaluation Questionnaire of the First Study**

Table 3 shows the results of the first question of the evaluation questionnaire phase which asked in which finance product the participant would invest her money after conversing with the system. Only 16% remained indifferent or undecided. There is no statistical difference between the answers of the single-bot and the multi-bot groups in terms of chosen financial product after the interaction. Also, we analyzed the conversations which had a recommendation for investment and compared with the choice of investment indicated in Q1. Of those, 82% of the single-bot and 86% of the multi-bot participants agreed with the advice provided but there was no statistically significant difference in taking advice from the single-bot and multi-bot versions.

Table 4 displays the results of questions Q2 to Q11 of the evaluation questionnaire. Participants seem to find both versions similarly informative, knowledgeable, and trustful as shown by the overwhelming agreeing with Q2, Q3, and Q9. However, in general they found both experiences inferior to speaking to a bank manager, as expected, questions Q6 and Q7. Participants seem to split in whether they would trust the system to make a decision (Q5) or change their current finance portfolio after the study experience (Q8). Participants also found the system innovative (Q10 and Q11) and different from any advice system they encountered before.

To compare the single- and multi-bot versions, a *Chi-Square Test* was applied to the distributions of each question, after grouping the numbers of participants who *disagree* with *partially disagree* and the *agree* with the *partially agree* fields to compensate for some low numbers and splitting the value of the *same* field evenly among the other two columns. The results show that no question of the evaluation questionnaire had results significant different in the single-bot and the multi-bot versions as shown in the rightmost column of table 4.

We did not find any influence of demographics, finance knowledge, and risk profile of the participants in each group and the way they answered the evaluation questionnaire. For the comparison between the groups of participants using the single-bot version and the multi-bot version, we used the *Chi-square test* of independence (*p<=0.05*) for each demographical characteristic such as gender, age range, interest in finance, and exposure to *Artificial Intelligence* technology and all the questions of the second questionnaire. The relation between those variables was not statistically significant (*chi-square > 0.05*), which suggests that characteristics of the participants of each group did not influence their experiences using the systems. Moreover, we found no difference between the participants' knowledge about finance and their risk attitudes (*chi-square > 0.05*). For this analysis, we considered that a participant had a risky attitude if she did not have an average or high knowledge of a given financial product and responded in the second questionnaire (Q1) that she would invest in a different product.

**Findings of the Analysis of the Conversation Logs of the First Study**

All the conversations during the free interaction phase were logged. Since participants could stop the experiment at any time, the duration and number of utterances varied considerably, but, as shown on table 2, basic statistics related to the length, duration, and average share of the conversation by chatbots and participants were very similar for the two groups.

Next, we explored whether there were differences on the number of *types of user intent* and their associated *categories* (as described before in the data analysis methodology) of the utterances made by the participants according to the version they interacted with. To make the analysis more consistent, we discarded the conversations where the participants made less than 5 or more than 25 utterances and excluded the answers to the initial simulation prompted by *InvestmentGuru*. This resulted in 19 and 22 conversations logs for single-bot and multi-bot, respectively.

Table 5: Types of user intent identified in the transcripts of the free-interaction phase.

| TYPE OF USER INTENT | SINGLE-BOT | | MULTI-BOT | | TOTAL |
|---|---|---|---|---|---|
| | number | % | number | % | number |
| core - cdb | 26 | 11% | 55 | 21% | 81 |
| core - savings | 18 | 7% | 16 | 5% | 32 |
| core - trea. bonds | 49 | 20% | 68 | 23% | 111 |
| simulation | 28 | 12% | 26 | 9% | 52 |
| comparison | 27 | 11% | 31 | 10% | 54 |
| generic finance | 36 | 15% | 35 | 11% | 64 |
| other investments | 8 | 3% | 30 | 1% | 9 |
| chitchat | 21 | 9% | 37 | 9% | 46 |
| convers. problems | 30 | 12% | 1 | 11% | 60 |
| total intents | 243 | 100% | 362 | 100% | 509 |

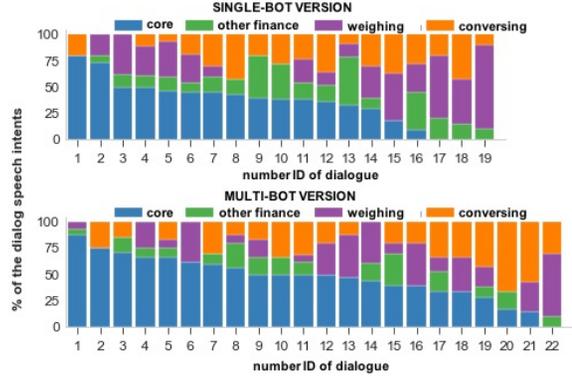

Figure 5: Proportion of the participant utterances for each category of user intent for single- and multi-bot.

The labeling of the conversation logs was performed by three of the authors of this paper. They were asked to classify the intent of each user utterance into the 10 types of user intent described before. The Fleiss's kappa coefficient $\kappa$ was used to assess the consensus amongst the annotators' labels. We obtained $\kappa = 0.84$, for the 769 user utterances (a value above 0.81 is normally considered almost a perfect level of agreement).

Table 5 shows the number of participant utterances for each type of user intent which resulted from the labeling process. Since all participants went through the initial simulation, led by *InvestmentGuru*, we did not include those utterances in the rest of the log analysis. To compare the remaining 9 types of utterances, we performed a Chi-test which determined that the distributions of types of user intent of single-bot and multi-bot were significantly different (*p-value=0.031 < 0.05*).

Figure 5 depicts the proportion of participant utterances of each of the 4 categories for the conversations in the single- and multi-bot versions for each conversation logged of each group of participants. The graphs suggest that participants in the single-bot version spent a smaller proportion of their conversations asking questions about the *core* products of *finch* but seem to do more utterances in the *other finance* category.

To check those visual insights, we calculated the proportion of the *core* utterances of each conversation, obtaining an average of 0.36 and 0.47 for the single- and multi-bot versions respectively. Comparing the two distributions with a *one-tailed T-Test*, we obtained that the multi-bot proportion of *core* utterances is greater than to the single-bot proportion (*p-value=0.045 < 0.05*). Similarly, we compared the proportion of *other finance* utterances, which yielded averages of 0.17 and 0.10 for the single- and multi-bot versions, respectively. The *one-tailed T-Test* indicated that the single-bot proportion of *other finance* utterances is greater than the multi-bot proportion (*p-value=0.023 < 0.05*). The other two categories, *weighing* and *conversing*, had single- and multi-bot averages of 0.26 and 0.19, and 0.22 and 0.23, respectively, but no statistically significant difference in the *weighing* or *conversing* categories was found between the single-bot and multi-bot versions.

### 4.4 Discussion of the Results of the First Study

The analysis of the evaluation questionnaire in the previous section shows that participants found both versions useful and knowledgeable, and that the majority would take the advice from the system. The analysis also showed no fundamental difference between the single-bot and the multi-bot groups in terms of their ability to support the participants to accomplish the proposed task of choosing a financial product. The kind of system participants interact with does not seem to influence which investment product they chose, since the distributions of answers to Q1 were indistinguishable statistically. This analysis provides initial evidence to answer positively research question RQ1 that the task outcomes are similar when collaborating with a personal chatbot or an equivalent multi-bot chat.

Questions Q2 and Q3 were designed to check typical elements associated with perception of competence: whether the system provided helpful information and had enough knowledge, respectively. The distributions of the participants' agreement with the statements were, as presented before, not statistically different. Similarly, Q4, Q6, and Q8 deal with the participants perception of how effective the system was, by exploring if they feel confident to use what they learn to advice someone else, use the system instead of talking to financial expert, or reconsider their investment options. As discussed in the findings section, the distributions of the two groups are very similar. Finally, Q5, Q7, and Q9 explore how much the participants trusted the systems they interacted with, by measuring their agreement with statements about trusting the system to make a decision, comparing it to the manager of their bank, and recommending it to a friend. For all purposes, participants in the single-bot and in the multi-bot version seem to trust in similar ways both versions as shown in the previous section.

Overall, the findings of the analysis of the evaluation questionnaire seem to support that the participants perception of competence, effectiveness, and trustworthiness is not affected by which system they interacted with. Those findings can be interpreted as providing evidence the task outcomes were similar for both versions because there was no significant difference in the way participants perceived key aspects (competence, effectiveness, trustworthiness) which could have affected how their final decisions were made. Of course, there are many other elements in play during a decision-making process which could be also assessed. We did not test those other elements for the sake of keeping the questionnaire in a manageable length. But since both the actual outcome and all those other elements did not yield significant differences, we believe it is safe to conclude that, from a task accomplishment and outcomes perspective, interacting with equivalent personal chatbots and multi-bot chats produced similar results in our experiments.

On the other hand, the analysis of the conversation logs seems to have found that participants behave in different ways when interacting with each version. The distribution of types of user intents was significantly different in the single-bot and multi-bot versions, indicating that the process of collaborating with a personal chatbot or an equivalent multi-bot chat do not seem to be similar, therefore providing evidence to refute RQ2. Here, Clark's theoretical framework of the joint-activity is important because it is founded on the idea that a collaboration process is indistinguishable from its associated language interchange: "Language use is really a form of joint action." [11]. Therefore, our finding that participants in each version asked questions of different types show not only that the conversation structure was different in each condition but seems to provide evidence that the collaboration process itself was different.

Further analysis on the proportions of different categories of participant utterances seem to provide some ideas to explain why the collaboration process was likely to be different in each version. As shown in the previous section, the proportions of the conversations dedicated to each user intent category provided that the utterances related to the core competency of the systems were more frequent in the multi-bot version than in the single-bot case, and less frequent for utterances on generic finance and other investments. In other words, participants seem to understand better the actual scope of the system (that the system only provides information about three financial products) in the multi-bot version than in the single-bot version. Considering that understanding what kind of questions a chatbot is able to answer is a fundamental part of predicting its behavior, the findings seem indicate that participants were more able to predict the behavior of the team of expert chatbots than the behavior of an equivalent personal chatbot. According to Klein et al. [32], the ability to predict the behavior of the other agents is a key element to enable better collaborations, and may have helped participants in the multi-bot version to offset likely higher collaboration costs due to the larger number of participants.

However, the analysis of the *conversing* category of utterances yielded no statistically significant difference between the two versions. This was an unexpected result, since in human groups performing joint activity there is a tendency for communicative actions to increase as the number of participants increase. We consider those results as an indicative, albeit partial, that maintaining common ground did not increase when collaborating with multiple chatbots. We take this result with care since the limitations of current conversational technologies often produce an excess of dialogue failures (in both versions) which could have masked other common-ground management issues. However, we have also to consider that *finch* implements some advanced aspects of turn-taking, such as topic-based selection of the next speaker, which greatly facilitate the flow of the conversation and the maintenance of common ground.

In summary, our first study indicated that interacting with personal chatbots or multi-bot chats seem to not affect the task outcomes but that the conversation structure (and thus the collaboration process) is somewhat different. While it seems reasonable that the division of knowledge among multiple agents may simplify predictability of behavior, the indication of not extra common ground management costs was somewhat surprising. To explore further the possible differences between the collaboration process of the two models, a second, qualitative study was done with the aim of understanding better the mental and cognitive processes at play.

## 5 SECOND STUDY: SUPERVISED COMPARISON OF THE TWO VERSIONS

To understand further the results of the first study, a second study was performed. The second study was conducted under the supervision of one of the authors of the paper. The focus of this study was to understand in depth the participants' actions while collaborating with a personal chatbot or an equivalent multi-bot chat and, particularly, to explore issues related to RQ2.

### 5.1 Methodology of Second Study

The second study consisted of 3 phases: invitation, experiencing one version, and comparison to the other version. In the *experiencing one version* phase, participants basically went through the first study, followed by a debrief session. In the *comparison to other the version* phase, participants interacted with the other version of *finch*, repeating some of the actions done in the experiencing phase and then were asked to compare the two versions. Participants were encouraged to think aloud during the study. A gift was given as a reward to go through the study. In more detail:

**Invitation:** participants were initially recruited by a snowball sample and were selected to reflect a similar demographics distribution to the participants of the first study. They received an invitation by e-mail to perform the study in the laboratory.

**Experiencing one version:** participants were welcomed at the laboratory with a brief explanation (the same as in the e-mail invitation of the first study). They signed a consent form and were asked to follow the same steps of the first study. The supervisor asked them to think aloud as they proceed and the importance of them sharing their thoughts. No interruptions or questions were made during this phase. After 5 minutes of free interaction, the supervisor asked them to type the word "evaluation" in the chat and they got prompted to answer the evaluation questionnaire survey. Participants were not told for how long they should interact, nor were they interrupted. Following the questionnaire, the participants were interviewed in a semi-structured way to debrief their experience. The goal was to understand the participant's perceptions of the version they had interacted with and their ability to identify how many interlocutors were there. The semi-structured interview consisted of seven guided, open questions:

- How many participants are in this conversation? Who are they?
- What do you think this system knows about?
- What kind of investments do you think this system knows about?
- What do you think this system does not know about?
- How did you unveil what this system knows and does not know about?
- What is the role of *InvestmentGuru*? Explain.
- What is the information this system provides which in your opinion helps you make good financial decisions? Give examples.
- How would you describe this system for someone who was looking for a tool to help in a financial decision?

**Comparison to the other version:** In this phase, the participants interacted with the other version of *finch*. The supervisor asked them to repeat at least three questions they have asked the system in the *experiencing one version* phase (the supervisor had taken notes of the questions to help participants to remember them). Next, participants answered debriefing questions to compare both versions. The guided questions for this phase were:

- How many participants are in this conversation? Who are they?
- What is the role of *InvestmentGuru*? Explain.

- How would you describe this system for someone who was looking for a tool to help in a financial decision?
- Which of the two versions would you recommend to a friend with a similar life and economical style as yours? Explain.
- Which version do you think knows more? Explain.

**Data Analysis**

Our goal in this second study was to unpack the collaboration process experienced by the participants. The thematic networks [4, 38] approach was used as an analytical tool to organize the categories obtained from that analysis of the material collected in the study. A coding scheme emerged from the data by applying an inspired grounded-theory approach [56] and was guided by a sequential analysis in the context in which it was produced. Codes were drawn from their frequency in the data and their perceived significance to answer our research questions. Transcripts were coded in parallel to the participants' sessions; therefore, no more sessions were held after data saturation was reached [19, 21]. Additionally, to the emerged codes, we integrated codes for attributes of joint activity [11, 31] such as predictability; directability; and common ground management, which are themes related to RQ2. Joint activity attributes also helped to organize the emerged codes. In the context of this paper, we focus our discussion only on the codes connected to RQ2. Conversation logs were analyzed together with audio transcripts to understand better the dialogue flow and failures during the experience of collaborating with a single or a multi-chatbot system. The analysis was performed by one of the authors of the paper.

## 5.2 Results of the Second Study

We had 16 participants performing the second study experiment in the laboratory. Half of the participants were assigned to interact with the single-bot version first and the other 8 started the experiment conversing with the multi-bot version. Gender was balanced in both cases. Participants were aged between 18 and 42 years old. Both groups were similar in their level of financial knowledge. The majority had low knowledge in treasury bonds and CDB, and high in savings, like the participants of the first study. The duration of the study ranged from 30 to 45 minutes. In the *experiencing one version* phase, most of the participants interacted with the agents for 5 minutes. Only two participants requested to end the interaction before 5 minutes, claiming they did not know what to ask anymore. Most of the participants in the single-bot version made between 6 to 10 utterances, while in the multi-bot version participants made between 7 to 12 utterances. For the analysis, we only considered the conversation logs of the experiencing phase of the study, and discarded comments during the interaction with the second version, since we observed effects of learning in the comparison phase. Because of that, we considered only comments about the two versions which happened in the debriefing of the comparison phase, ignoring the ones made when participants interacted with the second version of *finch*.

**Findings of the Analysis of the *Experiencing One Version* Phase of the Second Study**

All the participants were able to identify the number of agents in the conversation correctly. Participants in both versions valued the feature of doing simulations and comparisons and showing return-of-investment as one of the most important information the systems provided. As said by participant 15: *Comparison is very important. It was the most frequent thing it did here. [...] It evaluates my needs asking about duration and value. (P14)*

The role of *InvestmentGuru* was perceived as a financial adviser and informative in the single-bot version, and frequently participants in this version did not remember its name, showing that in some way the system and the agent were perceived to be the same. In the multi-bot version, *InvestmentGuru* was identified as a moderator, adviser, guide, observer, facilitator, and translator, one which intermediated the content the other chatbots knew. *So, it's like he's an observer: look, all those artificial intelligences [the bots]. They can pass information on themselves separated so that he [the InvestmentGuru] knows these details and can summarize to us. (P16)* Overall, the role of *InvestmentGuru* was perceived as valuable to establish the common ground of messages in both versions and to give interaction instructions which were frequently followed by participants, for instance, when it told them the use of *@chabot_name* to directly address a chatbot.

A typical behavior identified in the multi-bot version was a participant asking the same topic question for different bots: *What is CDB? (P10)* First, *CDBGuru* answered and the following question was *What is treasure? (P10)*, which *TDGuru* also answered. In the multi-bot version, participants asked more often about the three financial products, as they recognized them as associated to the bots: *I do not understand much about either CBD or treasury bonds, I understand more about savings. I understand the system speaks of these three [investments] because they are the three gurus I had here. I did not think asking about other types of investments (P12)*.

We observed that participants in the single-bot version asked more about other investments and generic finance terms. However, they asked more often anchored questions, questions linked to the previous utterances provided by *InvestmentGuru*, such as: *This varies according to the [interest] rate and the CDI (In)*; and then P9 asks: *What was the annual [interest] rate? (P9)* Participant 9 justifies the behavior: *When I asked about CDB, he did not answer, and from there I rewrote other information, the [interest] rate. (P9)*

An essential organizing theme which emerged from the analysis was how participants changed the course of the conversation. Cases of this theme were more discernible when dialogue failures happened in the conversation, for example when the system could not provide a satisfactory answer, often prompting the participants to change the theme of the conversation. In the single-bot version, participants usually changed the topic (e.g., rentability, risk, liquidity) or the product subject (e.g., CDB, treasure, savings) when a dialogue failure happened. However, in the multi-bot version participants changed topics and the product subject with more frequency and not only when dialogue failures occurred: *I'm thinking here to whom I'm going to ask now. (P8)*. Also, the interface design assisted users to direct questions connected to the core products of the system: *[...] since each one was in a different color I think it became easier to understand (P15)*.

In both versions, participants asked more about investments they had reported having low or medium knowledge of, that is, treasury bonds and CDB. Only three participants asked about all the three investments, but they had higher financial knowledge than the others and seem to be testing the chatbot answers to see whether they matched their knowledge about investments.

The conversation problems, as a consequence of dialogue failures, were perceived by participants as equivalent in both versions. Repairing common ground was made by participants by repeating the questions and paraphrasing. Paraphrasing was done by simplifying the utterance or making them more objective. For instance, participants substituted pronouns by the name of the financial products. Another repair action which was frequent in both versions was scrolling back the page. Participants in the single-bot version usually scrolled back to the *InvestmentGuru* initial statements which explained the scope of the system: *You can ask questions about definitions, risk, liquidity, or profitability to our gurus. (In)*. Meanwhile, participants in the multi-bot version scrolled back more often to the results of simulations and comparisons. Also, participants in the single-bot version seem to expect more a decision from the system. *It does not know how to suggest investments. I asked: What investment do you advise me? He [the single-bot] doesn't know. (P15)*. We also observed that, in the multi-bot version, participants, thinking aloud, more often compared the investment themselves to determine what would be the best investment considering the information provided by the different chatbots.

In the evaluation questionnaire, at the end of the experiencing phase, some participants were not able to choose one of the investment options. Two participants (P3 and P14) had a higher knowledge of finance than the other participants and act more towards validating each system answers than to actually making a decision. Another participant was not motivated by making an investment, since she only saves money at home because *when you withdraw the money from the bank, it takes some of that money because [the banks] collect taxes. (P6)*. Other two (P5 and P15) were not confident to choose any option, P15 because the system gave similar rentability range for all the choices. P5 did not find the system helped him to decide. For the other participants, deciding what type of investment is the best seemed to yield similar results, although the small number of participants does not enable us to state a definitive answer about that.

All the participants in the second study did not have previous experience with financial advisors although 11 participants had already conversed with single-bot systems in other domains. From those, 7 had bad experiences with single-bot chatbots. We did not find evidence that those participants preferred more one of the two versions.

**Findings of the Analysis of the *Comparison* Phase of the Second Study**

In the debriefing session of the comparison phase, the majority of the participants stated that both versions had the same level of knowledge. However, all but four participants would recommend the single-bot one to a friend, often justifying that the single-bot version was less complicated and more direct. *In the first [multi] I had to use the @ to talk (P10). This financial adviser [single] knows all about it. He will not let you go off. This guru is unique. I would indicate the second (single). He is more organized, cleaner, it solves everything by himself, he is already the last face. It's more direct. Here I do not need to point out the other one that induces me how to ask a question. (P13).* On the other hand, the four participants who recommended the multi-bot version mention the multi-bot version *had more details such as taxes (P8)* and *it was more robust (P2).* Participants perceived the multi-bot as more complex, but still, we did not see any significant change in their ability to accomplish the investment decision goal.

## 5.3 Discussion on the Results of the Second Study

The analysis of the second study showed that participants evaluated both versions as equivalent in their level of knowledge and usefulness, confirming the results of the first study. Interestingly, the qualitative analysis allowed us to understand better how the process of collaborating in each version was similar or not, assisting us to answer our second research question (RQ2).

The second study findings indicated that the process of collaborating was not similar, in agreement with the first study. The multi-bot version seems to induce participants to ask more about the three financial products. The single-bot version seems to induce participants to explore more the content and focus more on understanding details of each financial product. These observations suggest why in the first study participants asked more about core investments. In some ways, single-bot participants tended to explore more in-depth the content while multi-bot participants more commonly spread their questions to differently available investments, possibly stimulated by the visible presence of three different chatbots.

Conversations failures were also identified as a factor to change the course of the dialogue flow. Failures had a stronger effect changing the direction of the dialogue in the single-bot version, while in the multi-bot version participants explored more the scope of the system and changed the subject in spite of them. The availability of multiple chatbots seems to have stimulated participants to spread their questions in different directions. In spite of participants in both versions had experienced the similar number of dialogue failures, it was noted that the two versions afforded different ways to manage those failures and to maintain and repair common ground. Those expectations support the results of the first study and that the two versions may foster different conversation structures.

However, in the comparison phase most participants indicated a preference for the single-bot version, justifying their choice by perceiving the interaction process as less complex than the one with the multi-bot version. Since this preference seems to go against the other findings, we aim to perform further qualitative studies to understand and unveil other factors which might have affected their decisions and personal attitudes towards conversation decision-making with chatbots. In particular, we have to consider that many of them were exposed to single-bot systems before, so familiarity effects may have affected their opinions.

## 6 FINAL DISCUSSION

In this paper we explored the similarities and differences of working with conversational personal assistants vs. teams of expert agents. To do so, we built a multi-bot chat system to support financial decision-making called *finch* and used it in two studies, comparing its user experience with an equivalent personal chatbot version where the presence of the different chatbots was masked from the user. Using the two systems, we investigated whether both the users' ability to accomplish a task and the collaboration process itself were affected by having the multiple chatbots hidden or not.

### 6.1 Main Conclusions and Results

The analysis of the results of the first, between-subjects, unsupervised, quantitative, study with 69 participants yielded that the task outcomes were similar for both the personal chatbot and the multi-bot chat versions. Also, participants did not report any significant difference in how knowledgeable the systems were, how effective they were, and how much they trust them. However, the conversation log analysis portrayed that there were statistically significant differences in the distribution of the participants' intents of utterances in each version, indicating differences in how the interaction occurred in both versions. Moreover, we found that participants attained more to the core competencies of the system in the multi-bot version, suggesting that they were more able to understand and predict better what the system was about. No evidence of significant conversation or coordination overhead was found.

The second study started repeating the first under the supervision and encouraged the participants to think aloud. Following, the 16 participants were introduced to the other version, played with it, and asked to compare the two versions. The analysis of the second study found evidence of the greater predictability of the multi-bot version and that the users appreciated the ability to choose which expert chatbots should respond a question. However, some of them considered the multi-bot version to be more complex and difficult to use. We also saw different strategies when coping with dialog failures and a possible depth vs. breadth tradeoff between the single- and multi-bot systems.

What both studies seem to support is that working with a personal chatbot or an equivalent multi-bot chat are basically similar in terms of accomplishing the users' task and, at the same time, but foster different collaboration processes without adding significant extra coordination costs. Interestingly, previous work on single- vs. multi-bot conversational systems [10, 40] had also produced evidence that the conversation structure was different, although in those works task outcomes were not studied. Like those studies, we also found differences in the way participants converse with each type of system, but our results seem to indicate that those structural differences do not affect task accomplishment.

Our findings are intriguing because they do not align well with what is known about human collaboration and even with our own personal experience of working in groups. Teamworking tends to be harder to perform than working with a single individual, often with considerable extra coordination costs [5, 41, 42, 48]. At the same time, working together, especially in diverse and balanced teams, usually produces better quality results [60]. Assuming that the results of our studies can be generalized, what could be so different between teaming with people and machines that yields almost opposite collaboration experiences?

One possible explanation is that today's chatbots have still very limited conversation abilities which make them hard to collaborate with. As discussed before, the intention-action paradigm used by most chatbots, including the ones in our and previous studies, makes them very brittle and unable to handle conversational context well. In simpler terms, even state-of-art chatbots are extremely limited in their ability to collaborate using natural language, what tends to make the coordination costs, especially common ground maintenance, very high for the user. Therefore, actual differences in collaborating with personal chatbots vs. multi-bot chats may become buried in the noise of basic common ground repair. Those issues may be more pronounced in our study since previous research in this theme employed the *Wizard-of-Oz* technique [10, 40], not real chatbots like in our case, what most likely shielded the participants from many of the dialogue failures common on actual chatbots.

## 6.2 Limitations of the Studies

It is important to see that the two versions of the financial adviser system we used were, in fact, very similar. The main perceivable difference was the different colors used on the heading bubble of the expert chatbots, as shown in fig. 2 and fig. 4. Using almost exactly the same content was necessary to make the comparison fair, but more visual or implicit distinctions among chatbots in the multi-bot version might have affected the overall user experience to a greater extent. In teams of human experts, we normally have people with different traits and personalities, possibly making users more susceptible to distinct points of view. In future studies, we plan to look whether making the chatbots more distinct (faces, writing styles, or personalities) affects the results.

It was evident that participants noticed that they were interacting with multiple chatbots, since each of them introduced itself at the beginning of the conversation and, as mentioned before, we used different colors for them. It was also possible to see that participants progressively seem to create the new, multi-bot mental model as they interacted with the multi-bot system, both by the statements in the logs and also in the qualitative study. The participants often directed questions to each bot, based on their predictions about what the chatbot was able to answer. Participants justified that the multi-bot system was more complicated mostly because of actions such as having to directly address the expert bots (using @) or having to choose which bot to send a query. Nevertheless, their concern with complexity may reflect an extra cognitive overload, and we plan to investigate this in future experiments. Similarly, another issue which may have affected our studies is the novelty of conversing with multiple chatbots. Although chat apps have made common conversation with many people at the same time, interacting with multiple chatbots is a new experience even for users who are barely familiar with chatbots.

Also, participants had, on average, conversations of about 6 minutes and 11 utterances (see table 2). It is arguable whether the conversations and their perceptions would be different if the interactions had lasted more time. We are looking into ways to study longer interactions or, even better, sustained conversations. Indeed, we felt that participants in our study got better in interacting with the chatbots as the conversation progressed. However, *finch* was designed to provide objective and quick advice, one of the requirements coming out of its user-centered design process. We are thus looking into other scenarios and contexts where longer interactions make sense and, in those, fatigue due to extra, hidden coordination costs may become more pronounced.

Our study did not look into the participants' mental models for collaboration with more than one machine, but we did ask participants if they ever had a conversation with a machine before. All the participants in the qualitative study did not have previous experience with financial advisors although 11 participants had already conversed with single-bot systems in other domains. From those, 7 had a bad experience with single-chatbots. We did not find evidence that those participants preferred more one of the two versions based on their previous experience. Moreover, we explored their impressions on both systems only in the second half of the qualitative study where their previous experience testing first one of the systems might have affected their mental models of them (although the order of interacting with each system was counter-balanced).. Notwithstanding, we did not find any correlation between the order of interaction and their evaluation of complexity.

## 6.3 Choosing Between Personal Chatbots or Multi-bot Chats

If, as suggested by our studies, collaborating with a single chatbot or with a team of chatbots produces similar results and does not incur in significant extra coordination costs, using a personal assistant or a team of expert agents model becomes a designer's choice. As more and better technology to manage multi-bot chats (such as [7]) become available, other advantages and drawbacks of each model can be considered to determine which one is more applicable. In some cases, especially in domains composed of many specialized sub-fields, the team of expert agents model may be particularly appealing. For instance, a medical adviser system may gain from splitting the user interaction into multiple, illness-specialized medical chatbots covering specific areas of medicine and bringing to the chat only those experts considered relevant to the case. Also, if the suggested breadth vs. depth trade-off is warranted, designers may prefer multi-bot chats in situations where exploring multiple alternatives and comparison are the most important aspects of a task.

Choosing between the personal assistant and the team of expert agents models depends not only on the domain and context of the application but also on business, ethical, and moral considerations. First, it is important to understand that currently available digital personal assistants, such as *Alexa* and *Siri*, have still very limited skills. To increase those, the most common approach is regarding the personal chatbots as a front-end to manually built *skills* [26] which represent other services, making them, in fact, a *meta-chatbot* [9]. This works reasonably well, initially, but as time progresses, the issue of managing competing and complementary services creeps in. Alternatively, by relying on multiple expert chatbots provided by different organizations and people, transparently managed by the user, we can arguably create a much more scalable architecture and business model, similar to the *iPhone* ecosystem and the *App Store*.

There are also issues of trust and privacy. Suppose the user wants to order a pizza, should she delegate to the personal assistant the decision where it should get it from? How would the user know whether the personal assistant is not being paid to propose a specific provider, like the ad pages in *Google Search* [25]? Although it may be very comfortable to delegate decisions to a personal assistant, there must be transparency in the decision process to allow users to determine from whose behalf the personal assistant is actually taking actions for. It seems to us that a system where the user chooses the providers to interact it and have explicit representations of each of them tends to warrant more choice and freedom.

There are also political and social considerations at play here. Personal assistant models are likely to increase the concentration of power and data in the hands of a few, already dominant tech enterprises. Instead, doing things in a more "bazaar"-like approach [52] may foster a more distributed, bottom-up network of providers which has proven, in contexts such as open-source software, to be not only more equitable, but more efficient and innovative than monopolized systems.

Another line of criticism of personal assistants stems from the work on robot rights. Petersen [49] argues that "…having such intelligent creatures do our dirty work would simply be a new form of slavery." Similar issues concerning the ethics of servant machines have been discussed, among others, by Graaf [18], Coeckelbergh [13], and Levy [37], who warned that servant machines (like personal assistants) tend to desensitize us to the suffering of other human beings. It seems to us that, although any chatbot could be always regarded as a servant, collaborating with a multitude of chatbots feels much more an equitable proposition than being served by a single, butler-like personal assistant.

## 6.4 Future Work

As the technologies for conversational systems improve and chatbots become more able to perform teamwork, will the findings of our study still hold? That is an important but hard question to answer because it also depends on the likely increase of user familiarity with chatbot teams. What our studies seem to indicate is that a team of expert agents model does not suffer, as one would expect, from effects of larger collaboration costs than personal assistants in the context of current technologies. This frees developers and designers of today from one of the possible drawbacks in employing multi-bot chats as their interaction model.

Particularly, one of the key questions we could not answer properly with the studies reported here is whether the cognitive load was different in the two versions. Participants in the second study seem to perceive greater complexity in the multi-bot version, which could be a symptom of heavier cognitive requirements to manage multiple chatbots. Given the duration of the interactions in the studies, it is unlikely that cognitive fatigue would affect the results, but we would expect those effects to appear in longer conversations. Similarly, it is necessary to study further how different the mental models are in the two versions, and whether and how it affects the interaction experience.

We also plan to perform studies to explore further some of the issues and limitations discussed before. For instance, we want to investigate whether more distinctive appearances or behaviors for each expert chatbot affects the outcome or the collaboration process, and whether increasing the number of chatbots in the conversation exposes issues of cost of collaboration. Also, the chatbots in our experiment did not talk to each other or argue directly, which would portray a closer resemblance to an actual meeting with experts. Although the chatbot orchestration system we used enables this kind of inter-bot interaction, we did not use it because it would not be replicable in the personal chatbot version and therefore would impair our comparison studies. Nevertheless, this is a rich feature of multi-bot chats which we plan to explore in more detail in the future.